\documentclass[aps,prd,secnumarabic,superscriptaddress, amssymb, amsmath,nobibnotes,nofootinbib,11pt]{revtex4}
\usepackage{amsfonts,amsmath,hyperref,url, color}
\usepackage{eurosym}

\begin{document}

\title{%
Quantum ergosphere and brick wall entropy}
\author{M. Arzano}
\email{}
\affiliation{Dipartimento di Fisica ``E. Pancini'', Universit\`a di Napoli Federico II,
Via Cinthia, 80126 Fuorigrotta, Napoli, Italy}
\author{L. Brocki}
\affiliation{Institute for
Theoretical Physics University of Wroc\l{}aw, Pl.\ Maksa Borna 9,
Pl--50-204 Wroc\l{}aw, Poland}
\author{J. Kowalski-Glikman}
\email{jerzy.kowalski-glikman@ift.uni.wroc.pl}
\affiliation{Institute for
Theoretical Physics University of Wroc\l{}aw, Pl.\ Maksa Borna 9,
Pl--50-204 Wroc\l{}aw, Poland}
\affiliation{National Centre for Nuclear Research, Ho\.{z}a 69, 00-681 Warsaw, Poland}
\author{M. Letizia}
\email{mletizia@uwaterloo.ca}
\affiliation{Dept. of Applied Mathematics, University of Waterloo, Waterloo, ON N2L 3G1, Canada}
\affiliation{Perimeter Institute for Theoretical Physics, Waterloo, ON N2L 2Y5, Canada}
\author{J. Unger}
\affiliation{Institute for Theoretical Physics University of Wroc\l{}aw, Pl.\ Maksa Borna 9,
Pl--50-204 Wroc\l{}aw, Poland}

\date{\today}

\begin{abstract}
We revisit the ``brick wall" model for black hole entropy taking into account back-reaction effects on the horizon structure. We do so by adopting an evaporating metric in the quasi-static approximation in which departures from the standard Schwarzschild metric are governed by a small luminosity factor. One of the effects of the back-reaction is to create an ergosphere-like region which naturally tames the usual divergence in the calculation of the partition function of the field.  The black hole luminosity sets the width of such ``quantum ergosphere". We find a finite horizon contribution to the entropy which, for the luminosity associated to the Hawking flux, agrees remarkably well with the Bekenstein-Hawking relation.
 
\end{abstract}
\maketitle


The discovery that black holes carry an entropy proportional to their horizon area $A$ divided by the Planck length squared $L^2_p$ according to the celebrated Bekenstein-Hawking formula
\begin{equation}\label{BES}
S = \frac{A}{4 L^2_p}\,,
\end{equation}
is now more than forty years old \cite{Bekenstein:1973ur,Hawking:1974sw}. Despite the numerous derivations of the entropy-area relation \eqref{BES} existing in a variety of approaches to quantum gravity (see  \cite{Carlip:2008rk} for a comprehensive listing), the fundamental question concerning the nature of the degrees of freedom responsible for such entropy has not found yet a conclusive answer. Since due to quantum effects black holes radiate thermally  \cite{Hawking:1974sw}, one of the earliest attempts at addressing this question was to look at the quanta of a field in thermal equilibrium at the Hawking temperature near the horizon \cite{Zurek:1985gd, tHooft:1985vk} for the origin the Bekenstein-Hawking entropy. As it turns out the counting of modes of the field needed for deriving the thermodynamic partition function of the field yields a divergent result due to an infinite contribution coming from the black hole horizon. 't Hooft noticed that introducing a crude regulator by requiring the vanishing of the field at a small radial distance from the horizon one can obtain a finite horizon contribution to the entropy proportional to the area. Appropriately tuning the distance of such ``brick wall" from the horizon one can exactly reproduce the Bekenstein-Hawking formula \eqref{BES}. This result, albeit suggestive, replaces the question about the origin of Bekenstein-Hawking entropy with a question about the nature of the brick wall boundary. In \cite{Arzano:2013tsa} the authors suggested that back-reaction of the Hawking radiation can excite the quasinormal modes of the black hole thus effectively creating a ``wall" of oscillations in the geometry close to the horizon. The Bekenstein-Hawking entropy can be then seen as emerging from an interplay between the degrees of freedom of the geometry and those of the field.

In the present letter we adopt a philosophy similar to \cite{Arzano:2013tsa} and study the effect of back-reaction on the field propagating in the vicinity of the black hole horizon. We do this by replacing the usual Schwarzschild metric by a dynamic, ``evaporating" metric first proposed in \cite{Bardeen:1981zz}, in which the effects of back-reaction are parametrized by the luminosity of the radiating black hole. After solving the field equations in such metric we proceed to the usual mode counting for the field. The key feature of our model is that the small luminosity creates a ``quantum ergosphere", a region between the apparent horizon and the event horizon which effectively acts as a brick wall providing a finite horizon contribution to the entropy. As we show below, within the small luminosity and quasi-static approximations we use we are able to reproduce the Bekenstein-Hawking result within good accuracy.

Our starting point is the result by Bardeen \cite{Bardeen:1981zz} (see also \cite{York:1983fx} and \cite{York:1983zb}) that the metric of a spherically symmetric black hole slowly emitting Hawking radiation has the following form
\begin{equation}\label{1}
ds^2=-e^{2\psi}\left(1-\frac{2m G}{r}\right)dv^2+2e^{\psi}dvdr+r^2d\Omega,
\end{equation}
where $\psi$ and $m$ are functions of the advanced time $v$ and the radial coordinate $r$. For $m$ constant and $\psi=0$ this metric reduces to the Schwarzschild one, while for $\psi=0$ and $m=m(v)$ it becomes the Vaidya metric. Following \cite{York:1983fx} we define the mass of the black hole at a given time to be $M(v)=m(v,r=2mG)$ and its luminosity to be $L=-\frac{dM}{dv}$. In this letter we work in the regime of small luminosity $LG\ll 1$ and up to first order in perturbation theory so that we can write
\begin{equation}\label{2}
m(v,r)=  M(v) \simeq M_0 - Lv\,.
\end{equation}
We also assume that $v$ is small in the stage of the black hole evaporation process that we are interested in, so that ${v}/{2 M_0 G}\ll 1$. This quasi-static approximation means that we are only interested in time scales much shorter than the half-life of the black hole.

In what follows we will focus on the near-horizon features of the metric \eqref{1}. To this end we introduce a new ``comoving" radial coordinate $\rho = r - 2M G = r - 2M_0 G + 2Lv G$ and assume that $\rho$ is small, of the same order as $GLv$, so that in our computations we will only keep terms which are at most linear in $\rho$ and $L$. Further, we use the residual coordinate invariance to set $\psi(r=2M G)=0$, which in our approximation conveniently makes the function $\psi$ disappear from all the linearized expressions. Indeed
\begin{equation}\label{3}
  \psi(r) = \psi(r=2M G) + \left.\frac{\partial\psi}{\partial r}\right|_{r=2M G}\, \rho
\end{equation}
and it follows from Einstein equations that $\partial\psi/\partial r\sim L$ at $r=2MG$, so that the first term in \eqref{3} vanishes, while the second is of higher order and can be neglected. In terms of the comoving radial coordinate the metric near $r = 2MG$ takes the form
\begin{align} \label{com-metr}
ds^2
=& -\left(\frac{\rho}{\rho +2M G}+4L G\right) dv^2 + 2 dv d\rho + (\rho +2M G)^2 d\Omega.
\end{align}

The metric \eqref{1} has several horizon-like structures. We first consider the apparent horizon (AH), defined as the outermost trapped surface, i.e. the surface from which no light ray can move outwards. One characterizes this feature with the help of the expansion $\Theta$ of a congruence of geodesics, which describes the change in volume of a sphere of test particles on the geodesic in consideration, emanating from a point. The trapped surface is a boundary of the set of points for which $\Theta\leq 0$ and the apparent horizon is defined as a surface for which $\Theta = 0$.

We define (see, e.g.\ \cite{Poisson:2009pwt}) null geodesics by their tangent vectors $l^\mu, l_\mu l^\mu = 0$ and introduce an auxiliary null vector $\beta^\mu$ with normalization $\l_\mu\beta^\mu=-1$. This auxiliary vector is needed, because the subspace of vectors normal to the null tangent vectors $l^\mu$, which we are interested in, is two-dimensional and $\beta^\mu$ fixes the ambiguity in choosing this subspace.

The expansion is defined by the equation
\begin{align}\label{def1}
\Theta = l^{\mu}{}_{;\mu} +\beta^{\mu}l^{\nu}l_{\mu;\nu}\; ,
\end{align}
and using the null vectors
\begin{align}
l^{\mu} = (l^v,l^r)=\left(1, \frac 12\left(1-\frac{2M}{r}\right)\right),\; \beta^\mu = (0,-1),
\end{align}
yields in Bardeen coordinates $(r,v)$ and in comoving coordinates $(\rho,v)$, respectively,
\begin{equation}\label{expansion}
\Theta(r) = \frac{MG}{r^2}-\frac{1}{4MG},\quad \Theta(\rho) = \frac{MG}{(\rho+2MG)^2}-\frac{1}{4MG},
\end{equation}
which vanishes for $r_{AH}=2MG$, i.e. for  $\rho_{AH} = 0$. We thus see that the position of the apparent horizon is the same in both coordinates, which is not surprising since $\Theta$ is a scalar.

The second horizon-like structure is the ``timelike limit surface" (TLS) \cite{York:1983fx} , which is defined as the surface beyond which future directed timelike motion (a static observer) is not possible. The TLS is characterized by the condition
 \begin{align}
g(\partial_v, \partial_v)(\mbox{TLS}) =g_{vv}(\mbox{TLS}) \equiv  0,
\end{align}
which is satisfied at $r_{TLS}=2M(v)G $ and $\rho_{TLS} = -8ML G^2$ in the Bardeen and comoving coordinates, respectively. In the comoving coordinates the redshift, which is proportional to  $g_{vv}^{-1}$, is finite at the apparent horizon. On the other hand in Bardeen coordinates TLS and AH are identical and the redshift will diverge there. Thus we see that the TLS is a coordinate-dependent concept, and therefore it is of a very limited use.

Last but not least \cite{York:1983fx} gives a working definition of the event horizon (EH) based on the condition 
\begin{align}
\ddot r = 0 ,
\end{align}
i.e. it characterizes the EH as the surface imprisoning photons for times long compared to the dynamical scale $4M$ of the black hole. According to this definition the EH in the Bardeen and comoving coordinates lies at $r_{EH} = 2M G-8MLG^2$ or $\rho_{EH} = -8MLG^2$, identically in both coordinates. One should also notice that in comoving coordinates the EH coincides with the TLS.

Notice that since the apparent horizon is a surface separating the regions of spacetime that can and cannot communicate with infinity, we can consider it to be a natural boundary in the geometry of a slowly evaporating black hole, in particular when one has to define the boundary conditions for a field living outside the black hole. We will use this observation to shed new light on the brick wall calculation of 't Hooft \cite{tHooft:1984kcu} by including the contributions due to the backreaction, here modelled by a small luminosity $L$. The original result of 't Hooft is that the free energy of a scalar field living outside the Schwarzschild black hole has a horizon contribution given by
\begin{align}\label{glgfl}
F =- \frac{2 \pi^3}{45 h} \left(\frac{2MG}{\beta}\right)^4 + \ldots
\end{align}
where $\beta$ is the inverse temperature and $h$ a small cut-off parameter with dimension of length.  From \eqref{glgfl}, using standard manipulations, one can calculate the thermodynamic entropy associated to the field and the resulting contribution from the horizon term above is proportional to the area of the black hole thus qualitatively reproducing the Bekenstein-Hawking entropy-area relation. As we will see, the consequence of the finite luminosity is that the arbitrary cut-off $h$ is replaced by the distance between the event and apparent horizon and hence the \textit{quantum ergosphere} \cite{York:1983fx},  the region between the AH and the EH, plays the role of a physically motivated brick wall.

In order to proceed with the counting of modes of the field we start by solving the field equation in the vicinity of the black hole horizon in  the comoving coordinates introduced earlier. A massless scalar field $\phi$ in this geometry with metric \eqref{com-metr} obeys the Klein-Gordon equation
\begin{align}\label{kgg}
 &\left(4LG + \frac{\rho}{\rho+2MG} \right) \partial_{\rho}^2 \phi + 2 \frac{\partial_{\rho} \phi}{\rho+2MG}\left(1-\frac{MG}{\rho+2MG}+2LG \right) + 2 \partial_{\rho} \partial_v \phi +\nonumber \\
&\frac{2}{\rho+2MG} \partial_v \phi - \frac{l(l+1)}{(\rho+2MG)^2} \phi = 0  .
\end{align}
Since we are interested only in the contribution coming from the vicinity of the horizon in the case of small luminosity we assume $\rho/(M_0G) \ll 1$, $LG \ll 1$ and use the quasi-static approximation $\frac{v}{2M_0G} \ll 1$, this results into 
\begin{align}
\frac{2}{\rho + 2MG}\left(1-\frac{MG}{\rho+2MG}+2LG \right) \approx \frac{1}{2M_0G}\left(1 + 4LG\right),
\end{align}
meaning that in our approximation all the terms explicitly depending on the time $v$ drop out. The Klein-Gordon equation \eqref{kgg} can now be written as
\begin{align}\label{kg_approx}
\left(4LG + \frac{\rho}{2M_0G} \right) \partial_{\rho}^2 \phi +\frac{1+4LG}{2M_0G} \partial_{\rho}\phi &+ 2 \partial_{\rho} \partial_v \phi +\frac{1}{M_0G}\left(1-\frac{\rho}{2M_0G}\right) \partial_v \phi\nonumber\\ &-\frac{l(l+1)}{(2M_0G)^2}\left(1-\frac{\rho}{M_0G}\right) \phi =0.
\end{align}
Since, as a result of our approximation scheme,  eq.\ \eqref{kg_approx} is $v$-independent, i.e. we have a quasi-static situation, we can make use of the standard WKB ansatz
\begin{align}
\phi(\rho, v)= U(\rho) e^{-i\omega v}e^{i\int^{\rho} k(\rho') d \rho'} .
\end{align}
The real part of \eqref{kg_approx} now takes the following form
\begin{align}
\left(4LG+\dfrac{\rho}{ 2M_0G}\right)(U''-k^2U)+\frac{U'}{2M_0G}+2\omega k U
  -\dfrac{l(l+1)}{(2M_0G)^2}\left(1-\frac{\rho}{M_0G}\right)U = 0 \label{kgr}.
\end{align}
In the WKB approximation we assume that the amplitude $U(\rho)$ varies slowly compared to the wave number
\begin{align}\label{wkb}
\dfrac{U'}{U} \ll k, \hspace{10px}   \frac{U''}{U} \ll k^2,
\end{align}
and therefore \eqref{kgr} reads
\begin{equation}
-\left(4LG+\dfrac{\rho}{ 2M_0G}\right)k^2+2\omega k
-\dfrac{l(l+1)}{(2M_0G)^2}\left(1-\frac{\rho}{M_0G}\right) = 0,
\end{equation}
which can be solved for $k$ giving
\begin{align}\label{sol}
k^{\pm} \approx \frac{\omega \pm \sqrt{\omega^2 -\left(4LG+ \frac{\rho}{2M_0G} \right) \frac{l(l+1)}{(2M_0G)^2}}}{4LG+ \frac{\rho}{2M_0G}},
\end{align}
where again we neglected the terms which are of higher order in our approximation scheme. These two solutions correspond to in- and outgoing modes, respectively and can be used to calculate the thermodynamic entropy associated to the field. Indeed we can now count  the number of modes of the field and derive the statistical partition function from which all relevant thermodynamic quantities can be derived.

By approximating the sum over $l$ with an integral, the number of modes with frequency up to $\omega$ is given by
\begin{align}\label{gw}
g(\omega) = \int_0^{l_{max}} \nu(l,\omega)(2l+1)dl,
\end{align}
where $\nu(l,\omega)$ is the number of nodes in the mode with $(l,\omega)$ \cite{Li:2000rk}. Such quantity can be explicitly calculated by considering the modes $k^{\pm}$ constrained to a length $\Lambda$ (which acts as an infrared regulator) with periodic boundary condition
\begin{align}
\Lambda = \nu\frac{\lambda}{2}=\nu\frac{\pi}{k} \rightarrow \pi\nu = \Lambda  k,\quad k = \frac{2\pi}{\lambda},
\end{align}
where $\lambda$ is the wavelength of the mode. Since the modes are $\rho$-dependent the number of nodes is given by the following integral
\begin{align}
2\pi\nu(l,\omega) &= \int_0^{\Lambda} k^+d\rho+\int_{\Lambda}^0 k^-d\rho\nonumber\\
&= 2\int_0^{\Lambda} \frac{\sqrt{\omega^2 -\left(4LG+ \frac{\rho}{2M_0G} \right) \frac{l(l+1)}{(2M_0G)^2}}}{4LG+ \frac{\rho}{2M_0G}}d\rho,\label{nodes}
\end{align}
where we used eq. \eqref{sol}. The equation for the number of nodes above differs from the one obtained in \cite{Li:2000rk} in two aspects. First, due to the approximations we made there is no dependence on the advanced time $v$ and second, we do not have to introduce a cut-off close to the horizon, since the finite luminosity prevents the integrand from diverging at $\rho=0$. The integration with respect to $l$ in eq.\eqref{gw} is taken over those values for which the square root is real and yields
\begin{align}\label{gfull}
g(\omega) = \int_0^{\Lambda} \frac{2(2GM_0+\rho)^4 G^4\omega^3}{3\pi(8M_0L G^2+\rho(1+4LG))^2}d\rho.
\end{align}

In the original brick wall calculation \cite{tHooft:1984kcu} it is assumed that the scalar field, whose entropy we are going to compute, vanishes beyond the brick wall, situated at a small distance $h$ from the black hole horizon $r$, so that all the relevant integrals have the lower limit at $r+h$.  In the case of the Schwarzschild black hole considered in \cite{tHooft:1984kcu} the apparent and event horizon coincide, $r=r_{EH}=r_{AH}$, however in our case they are different and we must decide at which of the two we impose the scalar field boundary conditions. Our argument relies on the observation that for the existence of Hawking radiation it is only important to have a horizon that prohibits certain modes from reaching the observer \textit{momentarily}. To see what we mean by this consider the Unruh effect, whose existence is related to the presence of an accelerated horizon. This horizon exists because if the Unruh observer would accelerate uniformly forever some light rays could never reach him. However, the Unruh effect will occur also if the observer slows down later on and all light rays will reach him in a finite time span, i.e there is no global horizon. In our scenario the horizon momentarily separating the spacetime is the apparent horizon because the expansion $\theta$ of light rays vanishes there. Remembering that the apparent horizon corresponds to $\rho=0$ we conclude therefore that the integration over $\rho$ should be from $0$ to $\Lambda$, where $\Lambda$ is the infra-red cutoff introduced before, whose explicit value will not interest us here, since the expression for the area contribution to the entropy does not depend on it. The leading contributions of the integral in \eqref{gfull} are thus given by
\begin{align}\label{gom}
g(\omega) = \frac{4\omega^3G^2M_0^3}{3\pi L} + \frac{2\omega^3\Lambda^3}{9\pi(1+8GL)}\,,
\end{align}
where the second term is the usual volume contribution and has no relevance for our discussion. The thermodynamic partition function of the field is given by
\begin{align}
Z = e^{-\beta F} \,,
\end{align}
where $F$ is the free energy 
\begin{align}
\pi \beta F = \int dg(\omega) \ln{\left(1- e^{-\beta \omega}\right)}\,.
\end{align}
Using \eqref{gom} and neglecting the volume contribution to $g(\omega)$ we have
\begin{align}\label{ourf}
F = \frac{1}{\beta}\int_0^{\infty}\ln(1-e^{-\beta\omega})\frac{dg(\omega)}{d\omega}d\omega =-\frac{4M_0^3 G^2\pi^3}{45L\beta^4}\,,
\end{align}
from which we can calculate the entropy entropy of the field associated to the horizon boundary 
\begin{align}\label{entr}
S = \beta^2\frac{\partial}{\partial\beta}F = \frac{16M_0^3 G^2\pi^3}{45L\beta^3}.
\end{align}
Comparing our result for the free energy \eqref{ourf} with the standard result obtained from the brick wall calculation \eqref{glgfl} we see that the brick wall width parameter $h$ introduced by 't Hooft can be expressed in terms of the luminosity of the black hole as
\begin{align}
h =8 L M_0 G^2\,,
\end{align}
and thus the back-reaction of the quantum radiance on the horizon structures of the black hole naturally provides the regulator needed for a finite horizon contribution to the field entropy. 

In order to have an expression for the entropy \eqref{entr} to be compared to the Bekenstein-Hawking relation \eqref{BES} we now have to spell out the explicit form of the luminosity $L$ in terms of the black hole mass $M_0$.  In the first order approximation used in our calculation the luminosity $L$ is a small quantity so that we can identify it with the luminosity of Hawking radiation in the case of a Schwarzschild black hole. To find it, one considers \cite{fabbri2005modeling} a flux $X$ of radiation with energy $\omega_k$
\begin{align}
X(\omega_k) = \frac{\Gamma(\omega_k)}{2 \pi(e^{8 \pi M_0 G \omega_k} -1)}\,,
\end{align}
where the factor $\Gamma$ models the backscattering. Integrating the flux times the energy $\omega$ we find the luminosity that escapes to infinity
\begin{align}\label{lumin}
L = \int_0^{\infty} d \omega \thinspace\omega X(\omega).
\end{align}
The factor $\Gamma$ can be approximated by \cite{DEWITT1975295}
\begin{equation}
\Gamma \approx 27\pi M_0^2\omega^2
\end{equation}
and integration over $\omega$ yields
\begin{align}\label{lumi2}
L \approx \frac{1.69}{7680 \pi M_0^2 G^2}\;.
\end{align}
Plugging this expression in \eqref{entr} we finally obtain
\begin{align}
S =  0.79 \cdot 4 \pi M_0^2 G = 0.79 S_{BH}
\end{align}
where $S_{BH}$ is the Bekenstein-Hawking entropy. We thus see that our model does not reproduce the exact expression for the Bekenstein-Hawking entropy, but returns an expression correct within the $80\%$ accuracy, which is a remarkable result given the rather crude approximations that were used.\\

In this letter we showed how small back-reaction effects can be introduced in the derivation of the thermodynamic entropy of a field in thermal equilibrium in the proximity of a black hole horizon. The resulting changes due to a small but non-vanishing luminosity on the horizon structure of the black hole provides a natural brick wall regulator for the near-horizon modes of the field. Using the small luminosity and quasi-static approximations we were able to solve the field equations in the evaporating metric to find an explicit expression for the field modes, the degrees of freedom contributing to the thermodynamic partition function of the field. We showed that once the width of the quantum ergosphere is set by the Hawking luminosity the horizon contribution to the entropy of the field is in good agreement with the Bekenstein-Hawking relation for the black hole entropy. Given that in the original brick wall calculation the width of the brick wall had to be adjusted by hand in order to have the correct proportionality factor between entropy and black hole area, we find our result particularly suggestive given that in our model there is no arbitrary parameter that one has to tune to get the right result. Further work is needed to check whether better approximation schemes for the evaporating metric, for the field equations and their solutions can push us closer to an exact derivation of the Bekenstein-Hawking entropy from the degrees of freedom of a thermal field near the black hole horizon.

\section*{Acknowledgment} For LB, JKG, and JU this work is supported by funds provided by the National Science Center, projects number 2017/27/B/ST2/01902. ML acknowledges the Fondazione Angelo della Riccia and the Foundation BLANCEFLOR Boncompagni-Ludovisi, n\'{e}e Bildt for financial support. MA acknowledges support from the COST Action MP1405 ``QSpace" through a Short Term Scientific Mission Grant which funded a visit to the University of Wroclaw where part of this work was carried out.

\end{document}